\documentclass[lettersize,journal]{IEEEtran}
\usepackage{amsmath,amsfonts}
\usepackage{algorithmic}
\usepackage{array}
\usepackage[caption=false,font=normalsize,labelfont=sf,textfont=sf]{subfig}
\usepackage{textcomp}
\usepackage{stfloats}
\usepackage{url}
\usepackage{verbatim}
\usepackage{graphicx}
\usepackage{cite}
\usepackage{xcolor} 
\usepackage{multirow}
\usepackage{booktabs}
\usepackage{tcolorbox}
\usepackage[linesnumbered,ruled]{algorithm2e}
\usepackage{bbding}
\usepackage{adjustbox}

\usepackage{graphicx}
\usepackage{threeparttable}
\def\BibTeX{{\rm B\kern-.05em{\sc i\kern-.025em b}\kern-.08em
    T\kern-.1667em\lower.7ex\hbox{E}\kern-.125emX}}

\begin{document}

\title{Genetic Auto-prompt Learning for \\ Pre-trained Code Intelligence Language Models}

\author{{Chengzhe Feng, Yanan Sun, Ke Li, Pan Zhou, Jiancheng Lv, Aojun Lu}
\thanks{C. Feng, Y. Sun, J. Lv, and A. Lu are with the College of Computer Science, Sichuan University, Chengdu 610065, China (e-mail: fengchengzhe1@stu.scu.edu.cn; ysun@scu.edu.cn; lvjiancheng@scu.edu.cn; aojunlu@stu.scu.edu.cn).}
\thanks{K. Li is with the Department of Computer Science, University of Exeter, Exeter, EX4 4QF, UK (e-mail: k.li@exeter.ac.uk).}
\thanks{P. Zhou is with the Hubei Engineering Research Center on Big Data Security, School of Cyber Science and Engineering, Huazhong University of Science and Technology, Wuhan 430074, China (E-mail: panzhou@hust.edu.cn).}}

\markboth{}%
{}


\maketitle

\begin{abstract}
		As Pre-trained Language Models (PLMs), a popular approach for code intelligence, continue to grow in size, the computational cost of their usage has become prohibitively expensive. Prompt learning, a recent development in the field of natural language processing, emerges as a potential solution to address this challenge. In this paper, we investigate the effectiveness of prompt learning in code intelligence tasks. We unveil its reliance on manually designed prompts, which often require significant human effort and expertise. Moreover, we discover existing automatic prompt design methods are very limited to code intelligence tasks due to factors including gradient dependence, high computational demands, and limited applicability. To effectively address both issues, we propose {\bf Gen}etic {\bf A}uto {\bf P}rompt (GenAP), which utilizes an elaborate genetic algorithm to automatically design prompts. With GenAP, non-experts can effortlessly generate superior prompts compared to meticulously manual-designed ones. GenAP operates without the need for gradients or additional computational costs, rendering it gradient-free and cost-effective. Moreover, GenAP supports both understanding and generation types of code intelligence tasks, exhibiting great applicability. We conduct GenAP on three popular code intelligence PLMs with three canonical code intelligence tasks including defect prediction, code summarization, and code translation. The results suggest that GenAP can effectively automate the process of designing prompts. Specifically, GenAP outperforms all other methods across all three tasks (e.g. improving accuracy by an average of 2.13\% for defect prediction). To the best of our knowledge, GenAP is the first work to automatically design prompts for code intelligence PLMs.
\end{abstract}


\section{Introduction}
    \IEEEPARstart{C}{ode} intelligence is a key concern of software engineering, which
    leverages artificial intelligence to amplify software developer productivity. In recent years, a multitude of methods~\cite{allamanis2018survey} have been proposed to enhance code intelligence. Among those, the most popular method is to fine-tune Pre-trained Language Models (PLMs)~\cite{feng2020codebert,guo2020graphcodebert,ahmad2021unified,guo2022unixcoder} on a labeled dataset related to the specific downstream task. This is because vast amounts of general knowledge have been encoded within the PLM through pre-training on an extensive unlabeled text corpus. With fine-tuning, a PLM allows for the simultaneous utilization of both general knowledge and task-specific knowledge for downstream tasks.
	
     However, with the success of large PLMs~\cite{radford2019language,brown2020language} in deep learning recently, exemplified by ChatGPT, the size of PLMs has also grown significantly larger in code intelligence, inevitably resulting in increasingly expensive fine-tuning costs. For example, the recently proposed Code Llama model~\cite{roziere2023code} has a parameter size of up to 70B. Fine-tuning it would necessitate a minimum of 1680 GB of GPU memory. Clearly, this is not affordable to every researcher interested in the practice. Additionally, certain state-of-the-art PLMs are not customizable due to limited access to their weights, rendering the fine-tuning process impossible. An illustrative example is OpenAI's CodeX~\cite{chen2021evaluating}, a PLM preliminarily fine-tuned on publicly available code collected from GitHub. OpenAI has not open-sourced CodeX due to unknown reasons, and even more regrettable is the fact that they have also discontinued its API usage. 
     
     In the field of Natural Language Processing (NLP), there exist similar problems. To address the problems above, researchers in the NLP field propose prompt learning~\cite{ding2022openprompt}. Prompt learning enables PLMs to tackle downstream tasks (e.g., classification tasks) in the same way as their pre-training objective (e.g., masked language modeling~\cite{devlin2018bert}) with prompts. The prompt, consisting of a template and a verbalizer, is mainly carefully hand-crafted, thereby introducing human-prior knowledge for downstream tasks and PLMs. With the help of prompt, prompt learning can effectively stimulate PLMs to solve downstream tasks, achieving equivalent performance with fine-tuning. In addition, the computational cost of prompt learning is significantly lower than the fine-tuning method since prompt learning avoids model tuning process. Prompt learning has been proven effective in various NLP tasks. For instance, Sun \textit{et al.}~\cite{sun2022black} proved prompt learning can achieve equivalent performance and even outperform fine-tuning methods in sentiment analysis and other several real-world downstream tasks including topic classification, natural language inference, and paraphrase.

	Despite its success in NLP, the effectiveness of prompt learning in code intelligence is 
    obscured by the disparity between natural language and programming language. Hence, we first investigate the effectiveness of prompt learning for popular PLMs regarding three canonical code intelligence tasks, in particular the defect prediction~\cite{li2019improving}, code summarization~\cite{fernandes2018structured}, and code translation~\cite{nguyen2015divide}. To the best of our knowledge, this is the first investigation of prompt learning for code intelligence tasks. During the investigation, we found that a small change in the prompt can lead to a huge performance change in prompt learning for code intelligence tasks. For instance, only changing one token in the verbalizer can lead to a 6.66\% performance drop in the defect prediction task. Therefore, the prompt, including the prompt template and verbalizer, needs to be meticulously designed for a specific downstream code intelligence task. This design process, mainly through manual, requires substantial human effort and expert knowledge in both code intelligence tasks and related PLMs. Consequently, it has become a barrier to the wider adoption of PLMs for users who have no expertise in prompt design yet seek to utilize PLMs to solve code intelligence tasks. 
 
    To solve the aforementioned barrier, we first assess existing approaches for automatically designing prompts, originally proposed for resolving prompt design issues in NLP, to determine if they can substitute the manual prompt design process in code intelligence tasks. The majority of existing approaches are based on gradient descent (i.e., the soft prompt)~\cite{li2021prefix,liu2023gpt,vu2022spot,gu2022ppt}, which suffer from a lack of interpretability~\cite{hambardzumyan2021warp,khashabi2022prompt}, and are not applicable when gradients are inaccessible. Regrettably, the gradients of certain high-performance PLMs, such as CodeX, are inaccessible. Additionally, the approaches based on natural language tokens (i.e., the discrete prompt), introduce extra large models for assistance~\cite{prasad2023grips, xu2022gps} and targeted design for classification (understanding) tasks~\cite{hu2022knowledgeable,prasad2023grips}. Using discrete natural language tokens helps such methods avoid problems of gradient dependency and poor interpretability. However, it is worth noting that these approaches not only introduce high computational demands but also have limited applicability to certain code intelligence tasks. More unfortunately, we discover that these approaches often underperform in code intelligence tasks and very few existing approaches can design better prompts than manual design ones.
	
	To effectively address the above issues, we introduce {\bf Gen}etic {\bf A}uto {\bf P}rompt (GenAP), which can automatically design prompts based on a carefully tailored Genetic Algorithm~(GA)~\cite{forrest1996genetic}. GAs are powerful derivative-free optimization methods~\cite{kolda2003optimization}. By using bio-inspired operators such as mutation and crossover, GAs can generate high-quality solutions~\cite{mitchell1998introduction}. Commonly, the operators need to be carefully tailored for the particular problems at hand. For example, Sun \textit{et al.}~\cite{sun2020automatically} developed crossover and mutation operators for convolutional neural network architecture design. 
    Similarly, to apply GAs for prompt design, we develop crossover and mutation operators tailored specifically for this purpose. Leveraging the derivative-free nature of GAs, GenAP can design discrete prompts, thereby avoiding gradient dependency and ensuring interpretability. Moreover, GenAP also avoids those drawbacks introduced by discrete prompts. It can automatically design prompts for both understanding and generation tasks without introducing additional computational costs, except for the model inference cost alone. 
    
    To the best of our knowledge, GenAP is the first method specifically proposed to automate prompt design for code intelligence. Empirically, GenAP outperforms all other methods, including the manual prompt and other automatic prompt design methods from the NLP domain.
	
    In summary, the contributions of this paper are:
	\begin{itemize}
		\item We empirically investigate the effectiveness of prompt learning of PLMs for three canonical code intelligence tasks. The experiments indicate that the performance of prompt learning in code intelligence tasks is highly influenced by the design of the prompt used, underscoring the need for a careful prompt design process. However, this manual design process demands significant human effort and expertise, which impedes the widespread adoption of prompt learning. Thus we further analyze existing auto prompt design methods in the field of NLP and conduct an empirical study on their effectiveness in code intelligence tasks. We find that existing methods in the NLP field cannot be effortlessly applied to code intelligence tasks and very few existing methods can design better prompts than manual design ones.
		\item We propose GenAP, an automatic prompt design method specifically for code intelligence PLMs. GenAP leverages a GA to automatically design discrete prompts, thus avoiding the inherent drawbacks of soft prompt methods such as gradient dependency and lack of interpretability. GenAP requires no extra computational resources but only the inference demand of the used PLM. Moreover, GenAP supports both understanding and generation types of code intelligence tasks, exhibiting great applicability. With GenAP, we can effortlessly generate superior prompts compared to meticulously hand-crafted ones.
		\item We carefully tailor GAs for prompt design. GAs typically use fixed-length encoding strategies because the bio-inspired crossover operators are designed primarily for individuals of the same chromosome length. In these cases, the prompt length must be specified in advance. 
		However, the optimal prompt length is usually unknown beforehand. To address this, we propose a variable-length encoding strategy and corresponding crossover operators, enhancing the exploitation ability of GAs.
		\item We conduct experiments on three distinct code intelligence tasks with various code intelligence PLMs. Experiment results show that GenAP outperforms the performance of current auto prompt design methods in the NLP field across various experimental settings (e.g. improving accuracy by an average of 2.13\% for defect prediction). Notably, GenAP operates with the lowest computational cost, equivalent to the model's inference cost alone.
	\end{itemize}

    \begin{figure}[h]
		\includegraphics[width=\linewidth]{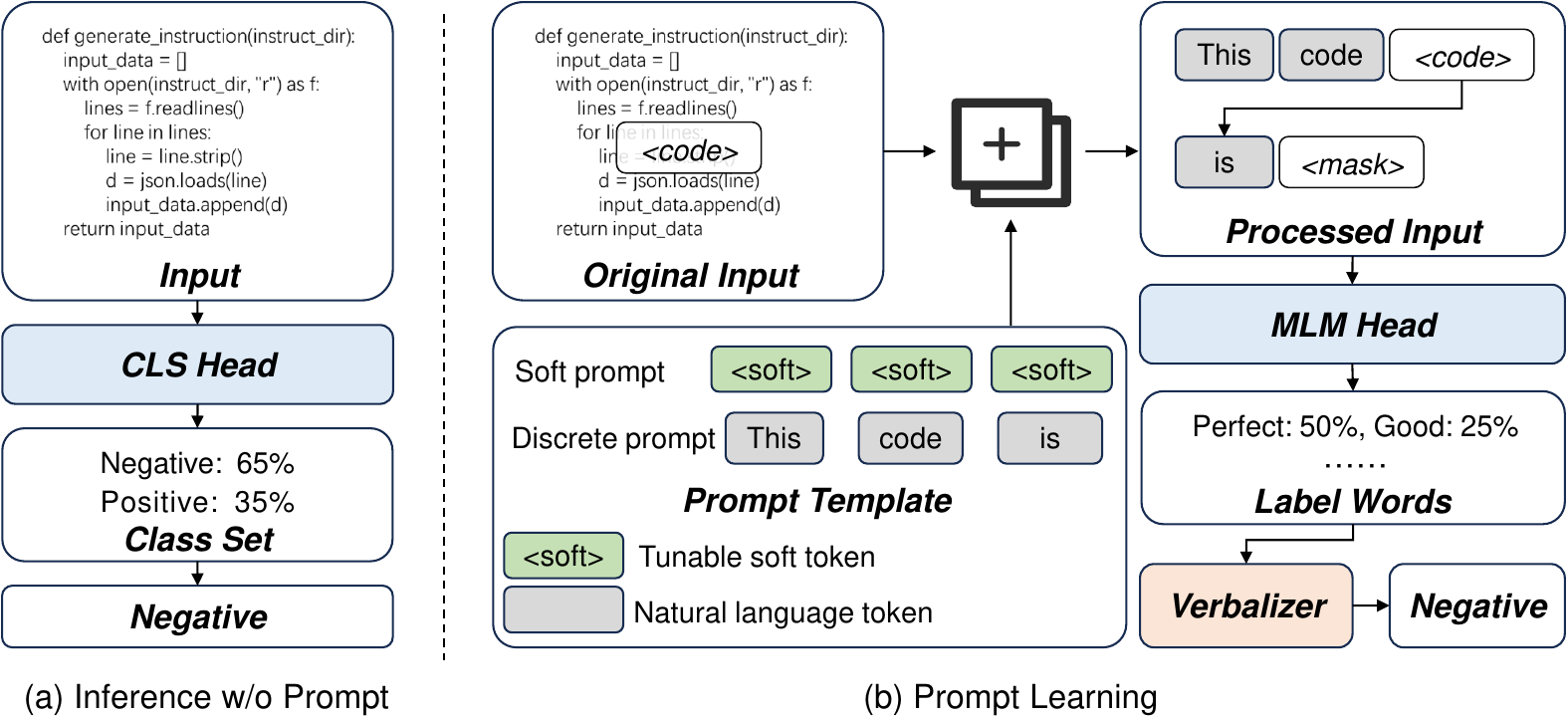}
		\caption{Illustration on the pipeline of Inference without Prompt~(a) and Prompt Learning (b), take defect prediction as an example.}
		\label{fig:fig1}
	\end{figure}

\section{Background}
    In this section, prompt learning and automatic prompt design, which are considered the background of the proposed method, are introduced to help readers better understand the proposed algorithm and the related works.
	\subsection{Prompt Learning}
	In this subsection, we mainly introduce the pipeline of prompt learning, the main components of the prompt, and two major types of prompts. 

    Initially, we introduce model inference without prompts to gain a better understanding of prompt learning. As illustrated in Fig.~\ref{fig:fig1}(a), when directly utilizing PLMs to solve the defect prediction task (i.e., model inference without prompts), the input code snippet will be processed by the CLS (Classification) head of PLM to obtain the class set, and the class with the largest probability will be the final prediction. In this way, there lies a gap between the pre-training objective (Masked Language Modeling~\cite{devlin2018bert}) and downstream tasks (e.g. Classification), where task-specific objectives and extra computation demands are introduced. Prompt learning, on the other hand, projects the downstream tasks to pre-training objectives for PLMs with the help of prompts. The prompt consists of a prompt template and verbalizer~\cite{schick2021exploiting,han2022ptr}. The prompt template is used to process the input text with some extra tokens, and the verbalizer is used to project original labels to words in the vocabulary for final prediction. 
 
     Assuming the prompt template is ``\textit{This code \textless code\textgreater\ is \textless mask\textgreater}", where the token \textit{\textless code\textgreater} stands for the original input code snippets and the token \textit{\textless mask\textgreater} is to be filled with predicted labels such as ``\textit{Perfect}". The final prediction is obtained by a verbalizer. We assume the verbalizer to be \{\textit{``Positive": ``Defective, Buggy", ``Negative": ``Perfect, Good"}\}, which maps the predicted labels ``\textit{Defective, Buggy}" to the final prediction ``\textit{Positive}", and 
     ``\textit{Perfect, Good}" to the final prediction ``\textit{Negative}". Now we have a code snippet as the original input. As shown in Fig.~\ref{fig:fig1}(b), this code snippet ``\textit{def generate\_instruction(instruct\_dir):\ \ldots}" will first be wrapped by a prompt template as ``\textit{This code def generate\_instruction(instruct\_dir):\ \ldots\ is \textless mask\textgreater}". This wrapped sentence will be sent to an MLM (Masked Language Modeling) head of PLMs to predict the ``\textit{\textless mask\textgreater}" token and then obtain the distribution of predicted label words ``\textit{Perfect: 50\%, Good: 25\%,\ \ldots}". The obtained distribution will be sent into a verbalizer to get the final prediction ``\textit{Negative}". 
	Based on the types of prompt templates, the prompt can be categorized into two types: Discrete Prompt and Soft Prompt. In the subsequent sections, we will provide detailed explanations of each prompt type.
	
	\subsubsection{Discrete Prompt} The discrete prompt~\cite{schick2021exploiting}, as shown in Fig.~\ref{fig:fig1}(b), is certain natural language tokens combined with the original input. Each natural language token is interpretable~\cite{gu2022ppt,liu2023pre}. For example, in the defect prediction task, natural language tokens ``\textit{This}" ``\textit{code}" ``\textit{is}", a code slot ``\textit{\textless code\textgreater}", an answer slot ``\textit{\textless mask\textgreater}" are combined to be a prompt template. This prompt template can be formulated as:
	\begin{equation}
		f_{prompt}(x) = \text{``\textit{This code \textless code\textgreater\ is \textless mask\textgreater}"}
	\end{equation}
	where ``\textit{\textless code\textgreater}" denotes the original input code snippet and ``\textit{\textless mask\textgreater}" should be substituted with labeled words from the verbalizer, such as \{\textit{``Positive": ``Defective", ``Negative": ``Perfect"}\}.

	\subsubsection{Soft Prompt} The soft prompt~\cite{lester2021power} can be seen as an alternative to the discrete prompt. Different from the discrete prompt, the soft prompt does not consist of natural language tokens but continuous vectors which can be learned during the tuning stage. As shown in Fig.~\ref{fig:fig1}(b), we call these continuous vectors ``tunable soft tokens", which are denoted as ``\textless soft\textgreater". Thus the designed prompt template can be formulated as:
	\begin{equation}
		f_{prompt}(x) = \text{``\textit{\textless soft\textgreater \textless soft\textgreater \textless code\textgreater \textless soft\textgreater \textless mask\textgreater}"}
	\end{equation}
	These tunable soft tokens are not human-interpretable since they are continuous vectors, making it difficult to map them back to the original discrete vocabulary. These soft tokens cannot be used when the gradients of PLMs are not accessible, since they are learned during the tuning stage through gradient descent. Furthermore, learned soft prompts cannot be reused across different PLMs unless they have the same embedding layer. This is due to the potential inconsistencies in vector representations caused by different embedding layers, leading to significant variations in the interpretation of learned soft prompts across different PLMs. 

    \begin{figure}[ht]
		\includegraphics[width=\linewidth]{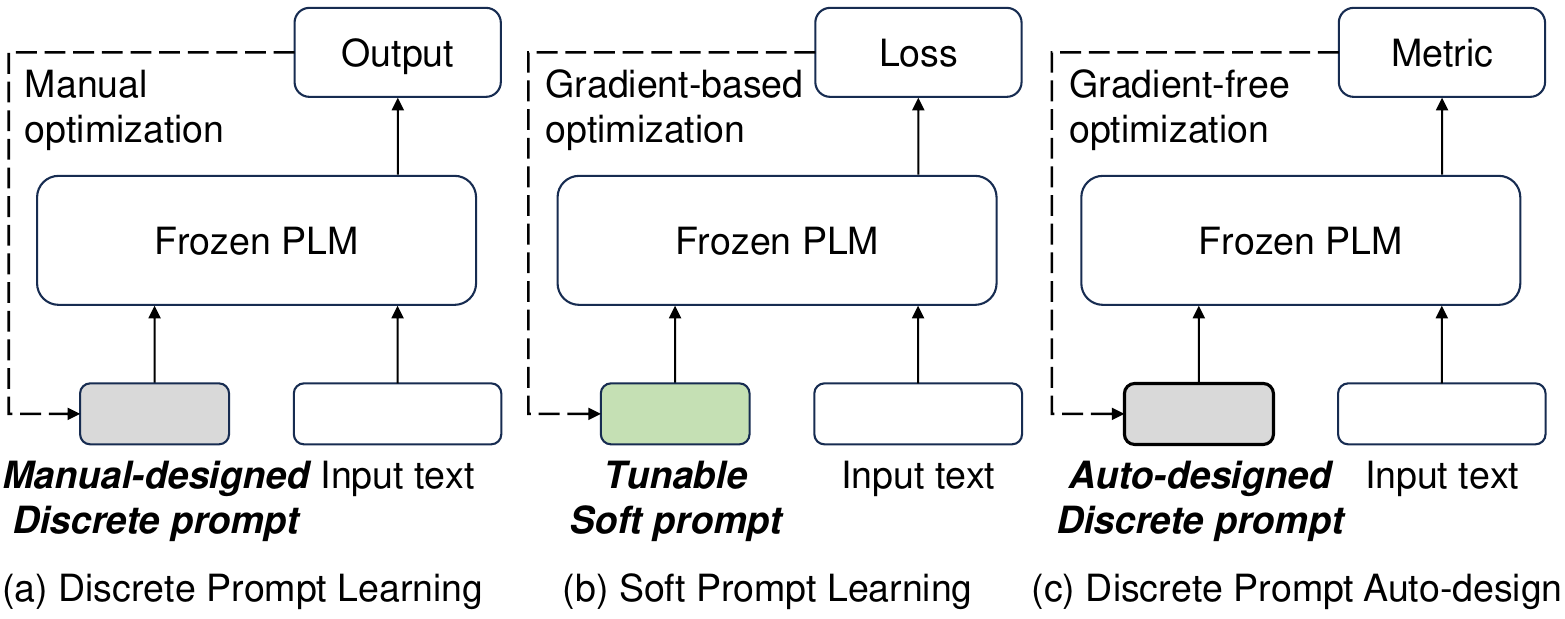}
		\caption{Illustration on the paradigms of Discrete Prompt Learning (a), Soft Prompt Learning (b), and Discrete Prompt Auto-design (c).}
		\label{fig:fig2}
	\end{figure}
 
	\subsection{Automatic Prompt Design}

    Discrete prompt learning, as depicted in Fig.~\ref{fig:fig2}(a), necessitates a labor-intensive manual optimization process. To avoid this time-consuming and labor-intensive manual optimization process, several automatic prompt design methods have emerged in the field of NLP. The majority of automatic prompt design methods, i.e., soft prompt learning Fig.~\ref{fig:fig2}(b), take advantage of soft prompts~\cite{qin2021learning,liu2023gpt,vu2022spot,gu2022ppt} since they can be learned through gradient-descent. Soft prompts introduce automation to these methods but also bring along a series of inherent drawbacks such as gradient dependency and poor interpretability. Considering that discrete prompts naturally lack the drawbacks of soft prompts, what they require is only a laborious manual design process. As a result, some researchers in NLP dived into discrete prompt auto-design Fig.~\ref{fig:fig2}(c). For instance, Xu \textit{et al.}~\cite{xu2022gps} propose GPS, which utilizes a genetic algorithm to automatically search for better prompts. Prasad \textit{et al.}~\cite{prasad2023grips} propose GrIPS, an automated procedure for improving prompts via an iterative, edit-based search. GenAP is also a discrete prompt auto-design method. 
	
	In this paper, we first analyze if the above methods can be directly applied to code intelligence tasks. In the case of methods that demonstrate potential for direct application to code intelligence, we proceed to evaluate and assess their performance specifically in code intelligence tasks.
	
	\section{Experimental Evaluation}
	
	\subsection{Research Questions} 
	In this section, we aim to answer the following research questions through an extensive analysis and experimental evaluation:
	
	\textbf{RQ1}: How effective is prompt learning in solving three canonical code intelligence tasks?
	
	\textbf{RQ2}: Which NLP auto-prompt design methods can be applied to code intelligence tasks, and how effective are those applicable methods when applied to code intelligence tasks?
	
	We design RQ1 to verify the hypothesis that prompt learning can boost the performance of PLMs on code intelligence tasks. RQ2 focuses on analyzing the applicability of existing automatic prompt design methods in the NLP field to code intelligence tasks. Additionally, it evaluates the effectiveness of those applicable methods when applied to code intelligence tasks.
	
	\subsection{Code Intelligence Tasks with Prompt Learning} 
	We evaluate prompt learning on three canonical code intelligence tasks, including one understanding task named defect prediction, and two generation tasks named code summarization and code translation. We describe the details of the chosen PLMs and the corresponding prompts for each task in the subsequent sections.
	
	\subsubsection{The Chosen PLMs} Following Wang \textit{et al.}~\cite{wang2022no}'s work on evaluating prompt tuning method, we choose CodeBERT and CodeT5 as the studied pre-trained models for evaluating prompt learning method.
	
	CodeBERT~\cite{feng2020codebert} is an encoder-only PLM that is based on RoBERTa model~\cite{liu2019roberta} and pre-trained on CodeSearchNet~\cite{husain2019codesearchnet} dataset. It is capable of encoding both source code and natural language text. CodeBERT has 125M parameters.
	
	CodeT5~\cite{wang2021codet5} is an encoder-decoder PLM built on T5 model~\cite{raffel2020exploring}. CodeT5 is pre-trained on the CodeSearchNet dataset with an additional C/C\# code corpus collected by its authors. It can solve both code understanding and code generation tasks. CodeT5 has two versions based on different parameter sizes, CodeT5-small with 60M parameters and CodeT5-base with 220M parameters. We choose CodeT5-base for the research endeavors since it has a larger parameter size. 
	
	
	
	
	
	
	
	\subsubsection{Defect Prediction Task and the Corresponding PLM} Given a code snippet, defect prediction~\cite{li2019improving,pradel2018deepbugs} aims to identify whether this code snippet contains defects that may be used to attack software systems, such as resource leaks, and DoS attacks. CodeBERT~\cite{feng2020codebert} is widely used to solve this task~\cite{kwon2022codebert,akimova2021survey}. Therefore, We choose it as the studied model for defect prediction. 
	
	\subsubsection{Code Summarization Task and the Corresponding PLM} Given a code snippet, code summarization~\cite{fernandes2018structured,hu2018summarizing} aims to generate a natural language comment to summarize the functionality of the code. Since CodeBERT has no decoder to generate comments, we utilize CodeT5~\cite{wang2021codet5} as the studied model for this task. 

    \subsubsection{Code Translation Task and the Corresponding PLM} Code translation~\cite{nguyen2015divide} aims at translating a code from one programming language to a different one. Since CodeBERT has no decoder to generate code snippets in a new programming language, we choose CodeT5~\cite{wang2021codet5} as the studied model for this task. 

  \begin{table}
		\caption{Statistics of the datasets used in this paper.}
		\label{tab:dataset}
        \begin{adjustbox}{width=\linewidth}
		\begin{tabular}{@{}c|c|c|ccc@{}}
			\toprule
			Tasks                                                                         & Datasets & Language & Train  & Valid & Test  \\
			\midrule
			defect prediction& Devign & C & 21854  & 2732  & 2732  \\
			\midrule
			\multirow{2}{*}{\begin{tabular}[c]{@{}c@{}}Code Summarization\end{tabular}} & \multirow{2}{*}{\begin{tabular}[c]{@{}c@{}}CodeSearchNet\end{tabular}} & Python   & 251820 & 13914 & 14918 \\
			& & Java     & 164923 & 5183  & 10955 \\
                \midrule
                Code Translation & CodeTrans & C\# \& Java & 10300 & 500 & 1000 \\
			\bottomrule
		\end{tabular}
        \end{adjustbox}
	\end{table}
	
	\subsection{Evaluation Datasets}
	
	To empirically evaluate the performance of prompt learning for code intelligence, we choose the datasets for the three tasks from the CodeXGLUE benchmark~\cite{lu2021codexglue}. CodeXGLUE is a popular code intelligence benchmark utilized by various works~\cite{guo2022unixcoder,wang2022no,wang2021codet5} for evaluation. Detailed information on the datasets, including the programming language used and dataset partition size, is illustrated in Table \ref{tab:dataset}.
	
	\subsubsection{Dataset for Defect Prediction} The dataset for defect prediction is the Devign dataset~\cite{zhou2019devign} from CodeXGLUE benchmark. It includes 27,318 manually labeled functions collected from two large C programming language open-source projects.
	
	\subsubsection{Dataset for Code Summarization} Follow CodeT5~\cite{wang2021codet5}, the dataset is the CodeSearchNet~\cite{husain2019codesearchnet} from CodeXGLUE benchmark, which contains thousands of code snippets and natural language description pairs for six programming languages including Python, Java, JavaScript, Ruby, Go, and PHP. We take its Java and Python parts as the used dataset. This is because these two programming languages share some similarities while also exhibiting distinct differences, making them suitable for the simulated experiment setting outlined in Section III-E.

    \subsubsection{Dataset for Code Translation} The dataset for code translation is from CodeXGLUE~\cite{lu2021codexglue} and is collected from four public repositories (Lucene, POI, JGit, and Antlr). This dataset provides Java (C\#) code and their corresponding translated C\# (Java) version.

	\subsection{Evaluation Metrics}
	In this subsection, we will introduce the evaluation metrics for three chosen tasks.
	\subsubsection{Evaluation Metric for Defect Prediction} For the defect prediction task, we follow~\cite{wang2022no} to use \textit{Accuracy} as evaluation metric. This metric is used to measure the model's ability to distinguish insecure code, which is formulated by Equation~(\ref{eq:acc}):
    \begin{equation}
    \begin{alignedat}{2}
    &Accuracy = \frac{ {\sum_{i=1}^{\left | D \right | } \theta_{i} } }{\left | D \right | }& \\
    &\theta_i = \begin{cases}
    1  \ \ \ if \ \ \ y_{i}=\hat{y_{i}}\\
    0  \ \ \ if \ \ \ y_{i}\neq\hat{y_{i}}
    \end{cases}&
    \end{alignedat}
    \label{eq:acc}
    \end{equation}
	where $D$ refers to the dataset and $\left | D \right |$ denotes its size. $y_{i}$ and $\hat{y_{i}}$ indicate the ground truth label and predicted label, respectively. 
	
	\subsubsection{Evaluation Metric for Code Summarization} For the code summarization task, we evaluate the quality of generated contents using the Bilingual Evaluation Understudy (BLEU)~\cite{papineni2002bleu} score, which is a commonly used metric in previous works~\cite{feng2020codebert,wang2021codet5,wang2022no}. This metric is formulated by Equation~(\ref{eq:BLEU}):
    \begin{equation}
    \begin{aligned}
    &\text{BLEU} = BP \times \exp\left( \sum_{i=1}^{n} w_{n}\log p_{n} \right) \\
    &\quad BP = \begin{cases}
    \begin{aligned}
    &1  &&if  \quad c > r\\
    &e^{1-r/c} &&if  \quad c \leq r 
    \end{aligned}
    \end{cases}
    \end{aligned}
    \label{eq:BLEU}
    \end{equation}
	In the experiments, aligning with~\cite{feng2020codebert,wang2021codet5,wang2022no} we employ a smoothed BLEU-4 score, which means $n=4$, to evaluate the generation tasks. In addition, $p_{n}$ refers to the modified n-gram precision and $w_{n}$ represents the weight assigned to it. Furthermore, $BP$ denotes the brevity penalty, and, $c$ and $r$ indicate the lengths of the generated comment and the target comment, respectively.
 
    \subsubsection{Evaluation Metrics for Code Translation} For the code translation task, we use two metrics, including the same smoothed BLEU-4 score as the code summarization task, and  CodeBLEU~\cite{ren2020codebleu}. CodeBLEU is a widely used metric in the code intelligence community for evaluating generated code contents~\cite{wang2022no,guo2022unixcoder,chakraborty2022natgen}. CodeBLEU can be defined by Equation~(\ref{eq:CodeBLEU}):
    \begin{equation}
	\begin{aligned}
        \rm{CodeBLEU} =& \alpha \cdot \rm{BLEU} + \beta \cdot \rm{BLEU_{weight}} \\
        +& \gamma \cdot \rm{Match_{ast}} + \delta \cdot \rm{ Match_{df}}
	\end{aligned}
    \label{eq:CodeBLEU}
    \end{equation}
    where $BLEU_{weight}$ is weighted n-gram matching score, $Match_{ast}$ is syntactic abstract syntax tree matching score, and $Match_{df}$ is semantic data flow matching score. In addition, $\alpha,\beta,\gamma,\delta$ serve as weights for each score, which are all set as 0.25 by following~\cite{wang2021codet5,lu2021codexglue}.
 
	\subsection{Implementation Details}
	
	\subsubsection{Experiment Settings} All experiments are implemented based on PyTorch and Huggingface Transformers. We implement prompt with OpenPrompt~\cite{ding2022openprompt} framework. We follow the parameter configuration described in~\cite{wang2022no} for the experiments in this paper. To ensure a fair comparison, all experiments are conducted with a fixed random seed. Following~\cite {wang2022no}, we set seed 52 for the defect prediction task and 42 for both the code summarization task and code translation task. We perform all the experiments on a server equipped with NVIDIA RTX 3090 GPU.
	
	\subsubsection{Baselines} In the prompt learning scenario, parameter tuning is avoided, and the PLMs are applied directly to the downstream task without being fine-tuned. To evaluate the effectiveness of prompt learning methods, we applied CodeBERT directly to the defect prediction task as the baseline without further fine-tuning. However, CodeT5's ability to code summarization is poor without fine-tuning, thereby insufficient to support experiments. Therefore, we first fine-tuned CodeT5 on a Python dataset to give it basic abilities and then applied it to a Java code summarization task to simulate direct inference. For code translation, the ability of CodeT5 is also poor without fine-tuning; thus, we fine-tuned CodeT5 on C\# to Java translation to test whether prompts can benefit a fine-tuned model or not.

        To ensure fair comparisons, all experiments involving CodeBERT and CodeT5 use the same configurations as described above. 

        \subsubsection{The Design of the Manual Prompt Template and Verbalizer} As is shown in Fig.~\ref{fig:fig2}, for the defect prediction task, the prompt learning methods require a prompt template and a verbalizer. To investigate the impact of different prompts, we follow the design of~\cite{wang2022no}'s design. For the code summarization task, the prompt template design is also based on the design of~\cite{wang2022no}, with the inclusion of a few additional prompt templates for experimental purposes. Note that there is no verbalizer for the generation task. For the code translation task, since the work in~\cite{wang2022no} does not release its design, we design several prompt templates for experimental purposes. Note that code translation is also a generation task thus there is no verbalizer for it. 
	
	\subsection{Experimental Results and Analysis}
        
	\subsubsection{RQ1: Effectiveness of Prompt Learning} 
	
	In this section, we study the effectiveness of prompt learning by comparing it with direct inference on the code intelligence tasks, i.e., defect prediction and code summarization. Additionally, we explore the impact of prompts on fine-tuned PLMs through code translation. The experimental results are shown below.

      \begin{table*}[!h]
		\caption{Classification accuracy of comparing the performance of CodeBERT model on defect prediction task via different prompt templates and verbalizers. The Baseline result is \textbf{bolded}.}
		\label{tab:pl_defect}
            \begin{adjustbox}{width=\textwidth}
		\begin{tabular}{@{}c|c|c@{}}
			\toprule
			Prompt template                                                        & Verbalizer                                                                           & Accuracy \\ \midrule
			\textit{The \textless code\textgreater\ code is \textless mask\textgreater.} & \{\textit{``negative": ``indefective", ``perfect", ``positive": ``bad", ``defective"}\} & 54.10\%  \\
			\textit{The \textless code\textgreater\ code is \textless mask\textgreater.} & \{\textit{``negative": ``no", ``positive": ``yes"}\}                                  & 53.40\%  \\
			\textit{The \textless code\textgreater\ code is \textless mask\textgreater.} & \{\textit{``negative": ``clean", ``perfect", ``positive": ``bad", ``defective"}\}       & 47.44\%  \\
			\textit{\textless code\textgreater\ The code is \textless mask\textgreater.} & \{\textit{``negative": ``clean", ``perfect", ``positive": ``bad", ``defective"}\}       & 45.86\%  \\
			\textit{\textless code\textgreater\ It is \textless mask\textgreater.}       & \{\textit{``negative": ``clean", ``perfect", ``positive": ``bad", ``defective"}\}       & 47.95\%  \\
		W/o prompt template                                                        & W/o verbalizer                                                                          & \textbf{45.86\%}  \\ \bottomrule
		\end{tabular}
            \end{adjustbox}
	\end{table*}

    \begin{table}[!t]
    \centering
    \caption{BLEU of comparing the performance of CodeBERT model on code summarization task via different prompt templates. The Baseline result is \textbf{bolded}.}
    \label{tab:pl_summarization}
    \begin{adjustbox}{width=0.85\linewidth}
    \begin{tabular}{@{}c|c@{}}
        			\toprule
        			Prompt template                                                                        & BLEU  \\ \midrule
        			\textit{Code \textless code\textgreater\ Summarization \textless mask\textgreater.}                   & 17.39 \\
        			\textit{\textless code\textgreater\ Summarization \textless mask\textgreater.}                       & 17.34 \\
        			\textit{Summarize \textless code\textgreater\ \textless mask\textgreater.}                         & 17.06 \\
        			\textit{Generate comments for java \textless code\textgreater\ \textless mask\textgreater.}               & 17.18 \\
        			\textit{Generate comments for \textless code\textgreater\ \textless mask\textgreater.}              & 17.17 \\
        			\textit{Generate summarization for \textless code\textgreater\ \textless mask\textgreater.}          & 16.70 \\
        			W/o prompt template                                                                         & \textbf{17.29} \\ \bottomrule
    \end{tabular}
    \end{adjustbox}
    \end{table}
    
    \begin{table}[!t]
    \centering
    \caption{BLEU and CodeBLEU of comparing the performance of CodeT5 model on code translation task via different prompt templates. The Baseline results are \textbf{bolded}.}
    \label{tab:pl_translation}
    \begin{adjustbox}{width=0.95\linewidth}
        \begin{tabular}{@{}c|cc@{}}
                \toprule
                Prompt template                                                                & BLEU     & CodeBLEU                \\ 
                \midrule
                \textit{Translate \textless code\textgreater\ into java \textless mask\textgreater.} & 79.82     &  84.88          \\
                \textit{Translate \textless code\textgreater\ \textless mask\textgreater.}           & 79.19      &84.78     \\
                \textit{Csharp \textless code\textgreater\ to java \textless mask\textgreater.}         & 79.68         &84.97           \\
                \textit{Csharp \textless code\textgreater\ java \textless mask\textgreater.}         & 79.68         & 85.03           \\
                \textit{\textless code\textgreater\ \textless mask\textgreater.}                     & 77.15      &84.09              \\
                W/o prompt template                                                                  & \textbf{79.50}  & \textbf{85.08}\\ 
                Prompt tuning~\cite{wang2022no} &          79.76     &84.39          \\
                Navie copy & 18.69 & - \\
                \bottomrule
        \end{tabular}
    \end{adjustbox}
    \end{table}

	\textbf{defect prediction.} Table \ref{tab:pl_defect} shows the results of defect prediction. It is clearly observed that the prompt design strongly influences the effectiveness of prompt learning. A minor change in prompt can result in a significant change in performance. For instance, employing the prompt template ``\textit{The \textless code\textgreater\ code is \textless mask\textgreater.}" with the verbalizer \textit{{``negative": ``indefective", ``perfect", ``positive": ``bad", ``defective"}} results in an 8.24\% improvement compared to direct inference with CodeBERT. Conversely, utilizing the prompt template ``\textit{The \textless code\textgreater\ code is \textless mask\textgreater.}" with the verbalizer \textit{{``negative": ``clean", ``perfect", ``positive": ``defective", ``bad"}} yields a mere improvement of  1.58\%. Clearly, the performance of prompts with the same template and different verbalizers varies a lot. Only changing one word in the verbalizer leads to a 6.66\% performance drop. Similarly, when prompts share the same verbalizer but differ in templates, their performance ranges from +0\% to +2.09\%. These results indicate that prompt learning can benefit model performance and the design of prompts is the key factor in prompt learning. 
	
	\textbf{Code Summarization.} Table \ref{tab:pl_summarization} illustrates the results of code summarization. We can observe the same phenomenon as defect prediction. The performance of prompt learning ranges from a decrease of -0.59 to an improvement of +0.10 compared to the baseline, based on different prompt template designs. Specifically, utilizing the prompt template ``\textit{Generate comments for \textless code\textgreater\ \textless mask\textgreater.}" achieves a 17.17 BLEU score. By just changing one word ``\textit{comments}" to ``\textit{summarization}", the updated prompt template yields only a 16.70 BLEU score, representing a decrease of 0.47 compared to the previous score. 

    \textbf{Code Translation.} Table \ref{tab:pl_translation} illustrates the results of code translation. It is evident from Table \ref{tab:pl_translation} that the performance of CodeT5 varies from -2.35 to +0.32 in the BLEU score and -0.99 to -0.05 in the CodeBLEU score in comparison to the baseline, depending on the particular prompt used. Interestingly, we observe that after adding the best prompt listed in the table, the performance of CodeT5 exceeds that of prompt tuning~\cite{wang2022no}, which introduces a prompt during the tuning stage. These observations serve as evidence for the effectiveness of prompt learning. That is, prompt learning can also enhance a fine-tuned PLM.

	Based on the performance evaluation of the above tasks, it is evident that prompt learning can unlock the potential of PLMs. However, the effectiveness of prompt learning is highly dependent on the design of the prompt itself. With different prompts used, there are 8.24\%, 3.99\%, and 3.36\% performance variations compared to the baseline on three tasks.

	\begin{tcolorbox}[colback=gray!20, colframe=gray!20, arc=1mm]
		\textbf{Answer to RQ1:} Prompt learning can enhance PLMs and its performance is sensitive to the prompt design. Based on the specific prompt used, there are 8.24\%, 3.99\%, and 3.36\% performance variations compared to the baseline on three canonical tasks.
	\end{tcolorbox}

 	\begin{table*}[!ht]
		\caption{Comparison of different prompting methods for PLMs, in terms of several desirable properties. Note that ``U\&G" represents understanding \& generation tasks}
		\label{tab:comparsion}
            \begin{adjustbox}{width=\textwidth}
		\begin{tabular}{@{}cccccc@{}}
			\toprule
			Methods       & Gradients-free              & Fully Automatic             & Low Computing Burden        & Interpretability            & \multicolumn{1}{l}{Support Both U\&G} \\ \midrule
			Manual Prompt~\cite{schick2021exploiting} & \Checkmark   & \XSolidBrush & \Checkmark   & \Checkmark   & \Checkmark             \\
			Soft Prompt~\cite{qin2021learning}  & \XSolidBrush & \Checkmark   & \XSolidBrush   & \XSolidBrush & \Checkmark             \\
			Soft Verbalizer~\cite{hambardzumyan2021warp}  & \XSolidBrush & \Checkmark   & \XSolidBrush   & \XSolidBrush & \XSolidBrush             \\
			Knowledgeable Verbalizer~\cite{hu2022knowledgeable}         & \Checkmark   & \Checkmark   & \Checkmark   & \Checkmark & \XSolidBrush             \\
			GrIPS~\cite{prasad2023grips}        & \Checkmark   & \XSolidBrush & \Checkmark   & \Checkmark   & \XSolidBrush           \\
			GPS~\cite{xu2022gps}         & \Checkmark   & \XSolidBrush & \XSolidBrush & \Checkmark   & \Checkmark             \\
			\textbf{GenAP (Ours)}     & \Checkmark   & \Checkmark   & \Checkmark   & \Checkmark   & \Checkmark             \\ \bottomrule
		\end{tabular}
        \end{adjustbox}
	\end{table*}
  	\begin{table*}[!ht]
        \centering
             \caption{Performance Evaluation: CodeBERT in defect prediction task with Accuracy metric, and CodeT5 in Code Summarization and Translation Tasks with BLEU and CodeBLEU metrics, using various prompt design methods. Note ``-" indicates that a particular method is not suitable for specific tasks. The best results are \textbf{bolded} and the second best results are \underline{underlined}, same in all rest tables.}
             \label{tab:related_work_comparision}
        \begin{adjustbox}{width=0.75\textwidth}
        \begin{threeparttable}
		\begin{tabular}{@{}ccccc@{}}
			\toprule
			Methods                   & Accuracy         & BLEU (Sum.) & BLEU (Trans.)    & CodeBLEU (Trans.)     \\ \midrule
			W/o Prompt                & 45.86\%          & 17.29  & 79.50    & \underline{85.08}    \\
			Manual Prompt~\cite{schick2021exploiting}            & 54.10\%          & \underline{17.39} & \underline{79.82} & 84.88        \\
			Soft Prompt~\cite{qin2021learning}              & 54.14\%          & 17.34   & -   & -    \\
   			Soft Verbalizer~\cite{hambardzumyan2021warp}          & \underline{55.05\%}          & -  & -  & -           \\
			Knowledgeable Verbalizer~\cite{hu2022knowledgeable} & 52.93\%          & -  & -  & -           \\
   			GrIPS~\cite{prasad2023grips}                    & 54.10\%          & -   & -   & -         \\
			GenAP (Ours)            & \textbf{56.19\%} & \textbf{17.45} & \textbf{79.85} & \textbf{85.17} \\ \bottomrule
		\end{tabular}
            \end{threeparttable}     
        \end{adjustbox}
	\end{table*}
	
	\subsubsection{RQ2: Evaluation on Existing Auto-prompt Methods} 
	
	As depicted in RQ1, the effectiveness of prompt learning relies on the design of prompts. Typically, the prompts are manually designed. However, manual prompt design is a labor-intensive and highly specialized process~\cite{mishra2022reframing,wang2022language}, necessitating the development of automatic prompt design methods. While some automatic prompt design methods already exist in the NLP field, it remains to be seen whether these methods can be seamlessly applied to code intelligence tasks or not. In the following, we will examine the suitability of existing methods.
	
	\textbf{Soft Prompt}~\cite{qin2021learning} is an automatic prompt template design method wherein the prompt template is learned efficiently in continuous space with gradient descent. Soft prompt is automatic, and when the gradient of PLMs is accessible, it can be easily applied to code intelligence tasks. Nevertheless, its utilization comes with an additional computational overhead owing to the requirement for tuning prompts. In addition, it becomes impracticable when the gradient accessibility is restricted.
	
	\textbf{Soft Verbalizer}~\cite{hambardzumyan2021warp} is an automatic verbalizer design method that utilizes gradient descent to find suitable verbalizers. It suffers a similar problem to the soft prompt, i.e., it cannot be used when the gradient of PLMs is inaccessible. Furthermore, it cannot be applied to the generation tasks which not need a verbalizer.
	
	\textbf{Knowledgable Verbalizer}~\cite{hu2022knowledgeable} presents an automatic method for verbalizer design. This is achieved through the expansion of the verbalizer's label word space using external knowledge bases, followed by a meticulous refinement of this augmented space utilizing the capabilities of the PLM. This enhanced verbalizer space is subsequently employed for predictions. Given its gradient-free nature, this strategy is not bothered by gradient accessibility. However, it cannot be adapted to generation tasks, which inherently do not necessitate the use of a verbalizer.
 
	\textbf{GPS}~\cite{xu2022gps} is a semi-automatic discrete prompt design method that utilizes a genetic algorithm with three prompt generation strategies to enhance performance. It still requires manual prompts as initialization, necessitating human involvement. Additionally, the prompt generation strategies employed by GPS demand significant computational resources due to their reliance on T5LM-XXL, a PLM with 11B parameters and a minimum requirement of mere 44 GB GPU memory, which surpasses the capabilities of most existing GPUs. Moreover, GPS is not suitable for designing verbalizers, limiting its application to code intelligence tasks. Unless the following three conditions are met:  manageable computational demands, provision of an initial manual prompt, and absence of the need for verbalization, GPS can be utilized to solve code intelligence tasks.
	
	\textbf{GrIPS}~\cite{prasad2023grips} is a gradient-free, edit-based semi-automatic discrete prompt design approach. It employs edit operations relying on PEGASUS~\cite{zhang2020pegasus}, having a parameter size of 568M. GrIPS requires a minimum of mere 2 GB GPU memory, which is affordable for most users interested. However, GrIPS still relies on a manual prompt for initialization. In addition, GrIPS runs on a meticulously designed function for classification tasks. As a result, it is unsuitable for those generation tasks, restricting its applicability. In summary, GrIPS is only suitable for classification code intelligence tasks when an initial manual prompt is provided.
	
	As is summarized in Table \ref{tab:comparsion}, existing auto-prompt design methods have distinct deficiencies when applied to code intelligence tasks. They are only applicable to specific code intelligence tasks that can accommodate these limitations. 
    
    Next, we will assess the efficacy of these existing methods in tasks suited to their capabilities. We choose model direct inference and manual prompt method on the defect prediction task and the code summarization task as the baseline. All those auto prompt design methods are initialized with manual prompts illustrated in Tables \ref{tab:pl_defect} and  \ref{tab:pl_summarization}. Table \ref{tab:related_work_comparision} illustrates the experimental results.
	
	In the defect prediction task, all methods utilizing prompts outperform the baseline without prompts by a minimum of 6\%. This further shows the effectiveness of prompt learning for code intelligence tasks. However, compared to the manual prompt baseline, only a few auto prompt design methods achieve superior performance. Specifically, both the soft prompt and GrIPS methods perform on par with the manual prompt, while the knowledgeable verbalizer exhibits a performance deficiency of 1.17\%. In addition, only the soft verbalizer method demonstrates a 0.95\% performance boost. 
	
	In the code summarization and code translation task, we observe a similar phenomenon where prompt learning can enhance performance. However, in the code summarization task, the soft prompt method experiences a performance decline when compared to the manual prompt method.

      \begin{figure*}[!t]
    \centering
		\includegraphics[width=0.98\textwidth]{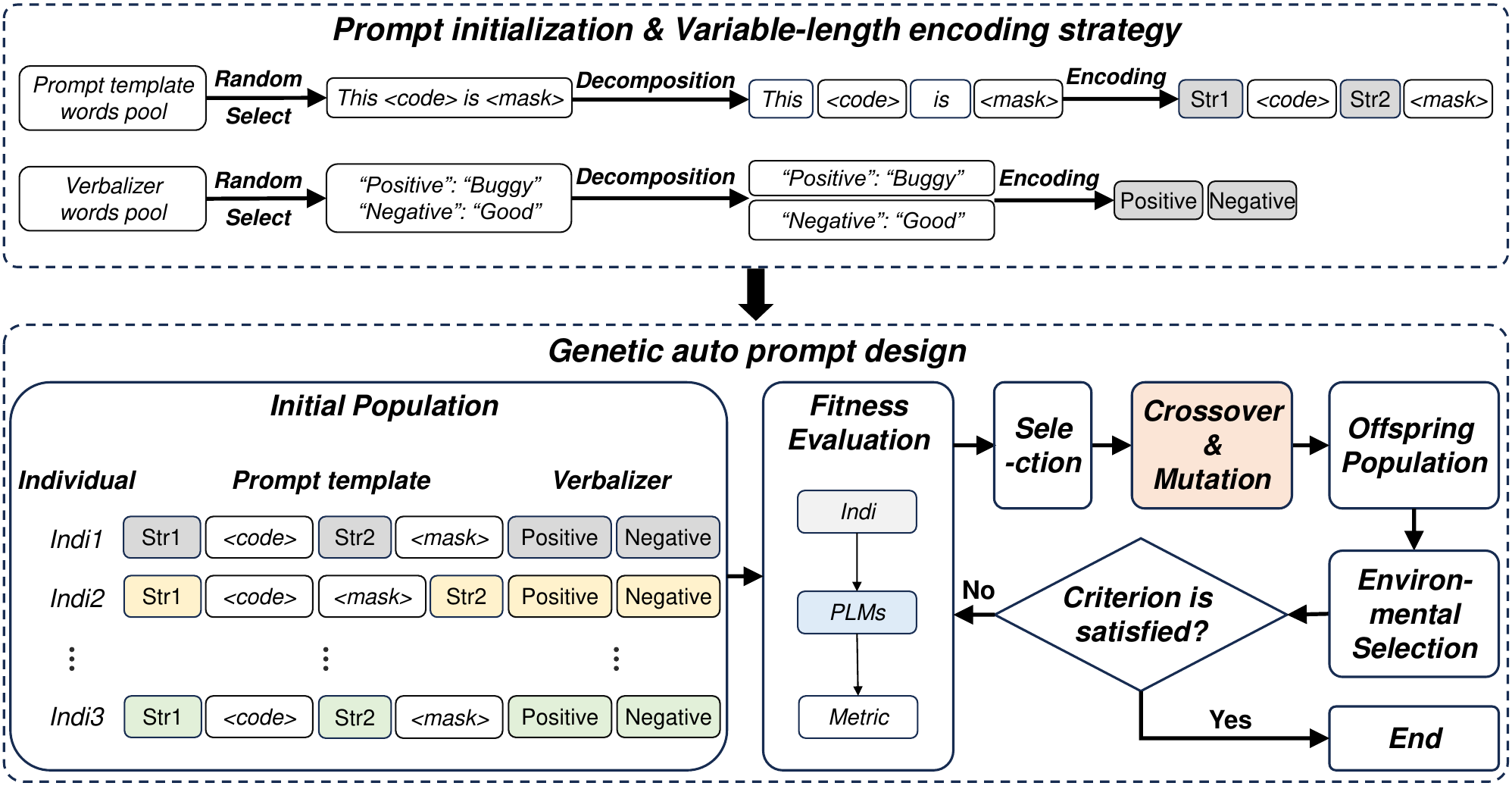}
		\caption{Illustration on the pipeline of GenAP, take defect prediction as an example}
		\label{fig:fig3}
	\end{figure*}

	\begin{tcolorbox}[colback=gray!20, colframe=gray!20, arc=1mm]
		\textbf{Answer to RQ2:} Existing automatic prompt design methods in the NLP field are only 
        applicable to specific code intelligence tasks that can tolerate their inherent limitations and constraints. Moreover, only a few existing automatic prompt design methods can enhance the performance in limited code intelligence tasks with distinct deficiencies (e.g. human efforts involved, gradients dependency).
	\end{tcolorbox}
	
	\section{Methodology}
	Our preliminary investigation above demonstrates the necessity of automatic prompt design methods and existing methods in the NLP field are not readily applicable to code intelligence tasks. To address this challenge, we propose Genetic Auto Prompt (GenAP). GenAP uses the Genetic Algorithm (GA)~\cite{forrest1996genetic}, which is a classical and powerful derivative-free optimization method~\cite{kolda2003optimization}, to automatically design discrete prompts. Please note that GAs serve as a comprehensive framework for addressing optimization problems by leveraging their inherent biological mechanisms. However, when implementing GAs in practical applications, it is crucial to tailor their components to the specific problems at hand. As shown in Fig.~\ref{fig:fig3}, we have meticulously designed the key components of GAs for GenAP, including prompt initialization strategy (i.e., prompt initialization \& variable-length encoding strategy), and crossover and mutation operators. In the following, we will provide a comprehensive explanation of GenAP, encompassing its mainframe, population initialization strategy, and crossover and mutation operators.
 	\begin{algorithm}[h]
		\SetKwInput{KwData}{Input}
		\SetKwInput{KwResult}{Output}
		\caption{Genetic Auto Prompt}\label{alg:one}
		\KwData{A set of predefined words to initialize prompt, an evaluation dataset, the population size, the maximal generation number}
		\KwResult{The discovered best prompt}
		$P_{0} \gets$ Initialize the population with the given population size using \textbf{the Prompt initialization \& Variable-length encoding strategy}\;\label{line1}
		$t\gets 0$\;\label{line2}
		\While{$t \textless  the\ maximal\ generation\ number$}{
			Evaluate the fitness of each individual in $P_{t}$ using acceleration components\;\label{line4}
			
			$Q_{t} \gets$ Generate offspring from the selected parent individuals using \textbf{the proposed mutation and the crossover operators}\;\label{line5}
			
			$P_{t+1} \gets$ Environmental selection from $P_{t} \cup Q_{t}$\;\label{line6}
			$t \gets t+1$\;\label{line7}
		}
		\textbf{Return} the individual having the best fitness in $P_{t}$.
	\end{algorithm}

	\subsection{Algorithm Overview}
	Algorithm \ref{alg:one} shows the mainframe of the proposed GenAP algorithm. Given a set of predefined words to initialize the prompt, an evaluation dataset, the population size, and the maximal generation number, the algorithm begins to work. During the evolution, a population is randomly initialized with the given population size using the proposed prompt initialization \& variable-length encoding strategy (line~\ref{line1}). Then, a counter $t$ indicating the current generation is initialized to zero (line~\ref{line2}). The key process of GenAP is to evolve the current generation iteratively. During the evolution, the fitness of each individual is evaluated on the given dataset (line~\ref{line4}). After that, the parent individuals are selected based on their fitness and then generated new offspring by the crossover and mutation operators (line~\ref{line5}). Then, the offspring population is obtained, and the environmental selection starts, after which a population of individuals surviving into the next generation is selected (line~\ref{line6}). Specifically, the current population consists of the parent population and the generated offspring population. Finally, the counter is increased by one (line~\ref{line7}), and the process continues until the counter exceeds the maximal generation.
	
	
	\subsection{Population Initialization Strategy}
    The proposed population initialization strategy consists of two strategies, i.e., the prompt initialization strategy and the variable-length encoding strategy. In this subsection, we will give a thorough introduction to the proposed prompt initialization \& variable-length encoding strategy.
	
	\subsubsection{Prompt Initialization strategy} 
	It focuses on generating prompts to initialize the population. We provide three methods for prompt generation tailored to different user needs. The first involves randomly selecting natural language words from the provided word pool to construct prompts. The second involves initializing prompts based on the input provided by users. The third combines prompts offered by users with randomly selected words from the provided word pool. These approaches allow users to obtain a prompt through fully automatic prompt design from scratch or semi-automatic prompt design by optimizing prompts provided by users. In the fully automatic prompt design scenario, we provide the word pool. In the semi-automatic prompt design scenario, users can either construct their own word pool or modify the one we provide. We construct the provided word pool for prompt templates by randomly extracting words from the vocabulary of CodeBERT without personal bias. The word pool for verbalizer is constructed of words describing positive or negative program characteristics. The rationale behind the construction process of the word pool will be thoroughly discussed in Section V.
	\subsubsection{Variable-length Encoding Strategy}
	This part of the algorithm aims to bridge the gap between typically used fixed-length encoding strategies by GAs and the unknown optimal prompt length. GAs typically use fixed-length encoding strategies because their bio-inspired crossover operators are designed primarily for individuals of the same chromosome length. In such scenarios, the prompt length must be predefined. However, as shown in previous experiments in Section III, optimal prompt lengths vary a lot. When faced with a new code intelligence task and PLM, it becomes nearly impossible to determine the optimal prompt length in advance. In order to retain the bio-inspired nature of GAs and utilize GAs to solve the prompt design problem, we design a method to transform the variable-length prompt into fixed-length encoding. As illustrated in Fig.~\ref{fig:fig3}, for the given prompt template \textit{``This \textless code\textgreater\ is \textless mask\textgreater"}, we divide it by recognizing \textit{``\textless code\textgreater"} as a division marker. Specifically, we project the words before the division marker into ``Str1" and the words after the division marker into ``Str2". Similarly, for the given verbalizer \textit{``Positive": ``Buggy", ``Negative": ``Good"}, we project the positive part of the verbalizer into ``Positive" and the negative part into ``Negative". This way, we manage to project prompts with various prompt templates and verbalizers into a fixed-length encoding.
	
	\subsection{Crossover \& Mutation Operators}
     In this subsection, we will provide a comprehensive introduction to the proposed crossover \& mutation operators. They are designed with a focus on the trade-off between computational cost and algorithmic effectiveness.
	\subsubsection{Crossover Operators}
	Crossover simulates the gene exchange process in biology. In GAs, it refers to selecting certain gene segments from two or more parent individuals and exchanging them to generate new offspring individuals. The purpose of crossover is to combine the strengths of different individuals, thereby increasing the diversity of offspring and potentially producing better solutions. Through crossover, one can avoid getting trapped in local optima and accelerate the convergence speed of the algorithm. 
	
	We design the crossover method in the light of the one-point crossover~\cite{srinivas1994genetic} in traditional GAs. The one-point crossover was originally introduced to operate on two individuals encoded with fixed-length encoding strategies. In this work, with the projection of variable-length prompts into fixed-length encoding in the proposed algorithm, the one-point crossover suits well. Thus, we have developed three types of crossover operators based on the one-point crossover. To illustrate, we consider two individuals in defect prediction tasks, denoted as \textit{indi1} and \textit{indi2}. The prompt template for \textit{indi1} is \textit{``This \textless code\textgreater\ is \textless mask\textgreater"} and for \textit{indi2} is \textit{``The code \textless code\textgreater\ works \textless mask\textgreater"}. Additionally, their verbalizers are as follows: ``\textit{Positive: Buggy, Negative: Good}" for \textit{indi1}, and ``\textit{Positive: Defective. Bad, Negative: Great}" for \textit{indi2}, respectively.
	
	\textbf{Type 1.} The Type 1 crossover operator partially exchanges the prompt template of \textit{indi1} and \textit{indi2}, specifically, exchanging the ``Str1" of two individuals or the ``Str2" of two individuals. Assuming an exchange of ``Str1", the prompt template of \textit{new\_indi1} and \textit{new\_indi2} will be \textit{``The code \textless code\textgreater\ is \textless mask\textgreater"} and \textit{``This \textless code\textgreater\ works \textless mask\textgreater"}, respectively.
	
	\textbf{Type 2.} The Type 2 crossover operator partly exchanges the verbalizer of  \textit{indi1} and \textit{indi2}. It is designed for tasks that require a verbalizer, and it will be deactivated for tasks that do not need a verbalizer. Specifically, it exchanges the ``Positive" of two individuals or the ``Negative" of two individuals. Assuming an exchange of ``Positive", the verbalizer of \textit{new\_indi1} and \textit{new\_indi2} will be ``\textit{Positive: Defective: Bad, Negative: Good}" and ``\textit{Positive: Buggy, Negative: Great}", respectively.
	
	\textbf{Type 3.} The Type 3 crossover operator wholly exchanges the prompt template of \textit{indi1} and \textit{indi2}.  After applying Type 3, the prompt template of \textit{new\_indi1} and \textit{new\_indi2} will be  \textit{``The code \textless code\textgreater\ works \textless mask\textgreater"} and \textit{``This \textless code\textgreater\ is \textless mask\textgreater"}, respectively.

    The three types of crossover operators will be randomly selected with equal probability during algorithm runtime. Despite the simplicity of the designed crossover operator, it effectively enhances the performance in discovering prompt designs, which will be empirically proved in Section V-D.
 
	\subsubsection{Mutation Operators}
	In GAs, mutation refers to randomly changing the values of specific genes within an individual to create a new individual. The purpose of mutation is to introduce new genetic information, enhance the exploration capability of the search space, and help the algorithm avoid getting trapped in local optima. Based on this concept, we design five mutation operators as follows:
	
	\textbf{Type 1.} The Type 1 mutation operator randomly removes several words from ``Str1" or ``Str2" or ``both Str1 and Str2" of the prompt template. The removed words are then inserted as a whole into the word pool to expand the list of candidate words.
	
	\textbf{Type 2.} The Type 2 mutation operator randomly selects one from the word pool and inserts it into ``Str1" or ``Str2" of the prompt template. Please note that a single run of the Type 2 mutation operator may not add just one word, as the Type 1 mutation operator can insert several words as a whole into the word pool. We design such an operator with the purpose of shortening the length of the prompt template, as the prompt template consumes valuable input length in PLMs.
	
	\textbf{Type 3.} The Type 3 mutation operator rearranges the order of ``Str1", ``Str2", \textit{``\textless code\textgreater"}, and \textit{``\textless mask\textgreater"}. For example, take ``This \textless code\textgreater\ is \textless mask\textgreater" as input, after applying the Type 3 operator, it could change to ``This is \textless code\textgreater\ \textless mask\textgreater". Then, it will be encoded again using the variable-length encoding strategy.
	
	\textbf{Type 4.} The Type 4 mutation operator randomly removes one word from the ``Positive" or the ``Negative" of the verbalizer. It is designed for tasks that require a verbalizer, and it will be deactivated for tasks that do not need one. 
	
	\textbf{Type 5.} The Type 5 mutation operator randomly selects one word from the word pool and inserts it into the ``Positive" or the ``Negative" of the verbalizer. It is also designed for tasks that require a verbalizer; for tasks that do not need one, it will be deactivated.

    These five types of mutation operators will also be randomly selected with equal probability during algorithm runtime. In addition, their effectiveness in improving the performance of discovering prompt designs will be empirically proved in Section V-D.
	
	\section{Implementation and Evaluation}
	
	In this section, we systematically evaluate the performance of GenAP in addressing five key research questions. Note that we maintain consistency with the datasets, models, and configurations used in Section 3 throughout this evaluation. We additionally introduce a larger PLM called CodeT5+~\cite{wang2023codet5+}. CodeT5+ is a family of code Large Language Models (LLMs). We choose its 770M parameter size version, the most popular version in the Hugging Face community, as the studied model to test the scalability of GenAP to LLMs. Although CodeT5+ surpasses lots of famous code LLMs like CodeX~\cite{chen2021evaluating} and CodeGeeX~\cite{zheng2023codegeex}, its ability in code summarization is poor without fine-tuning (similar to CodeT5), thereby insufficient to support experiments. Thus, we first fine-tuned CodeT5+ on a Python dataset for one epoch to give it basic abilities and then applied it to a Java code summarization task to simulate direct inference. For code translation, the ability of CodeT5+ is also poor without fine-tuning, thus we fine-tuned CodeT5+ on C\# to Java translation to test whether GenAP can benefit a fine-tuned model or not.
        
    As mentioned above, GenAP requires a word pool for initialization. To show the full automation ability of GenAP, the word pool in the following experiments is constructed in such a way with as little expertise and labor as possible. The word pool for the prompt template is not hand-crafted but randomly extracted from the vocabulary of CodeBERT without bias. We intentionally design such a seemingly suboptimal and ad-hoc word pool to reflect real-world scenarios where access to the vocabulary of PLMs may be limited. The word pool for the verbalizer is collected by searching for words describing positive or negative program characteristics and not cherry-picked.
    
	\begin{table}[!h]
        \centering
		\caption{Hyperparameter settings.}
		\label{tab:hp}
        \begin{adjustbox}{width=0.95\linewidth}
		\begin{tabular}{@{}c|c|c|c@{}}
			\toprule
			Hyperparameter  & Value & Hyperparameter                                                                                               & Value                                                                                                    \\ \midrule
			Population size & 20    & Max prompt length & 5 \\
			Crossover probability & 0.9   &   Mutation probability  & 0.4                                                                                                                                               \\ \bottomrule
		\end{tabular}
        \end{adjustbox}
	\end{table}
	
    The hyperparameter settings of GenAP are presented in Table \ref{tab:hp}. We adopt the hyperparameter setting following the GAs configuration presented in~\cite{sun2020automatically}, except for the mutation probability and max prompt length. The original setting of mutation probability in~\cite{sun2020automatically} is 20\%, but in this study, we have adjusted it to 40\%. The rationale behind this adjustment will be discussed in subsection V-C. Max prompt length is not a hyperparameter in~\cite{sun2020automatically} as their algorithm isn't designed for prompts. We set it to 5 to ensure a fair comparison with manual prompts.
 
	\subsection{RQ1: How Is the Effectiveness of GenAP Among Different PLMs and Downstream Tasks?} To answer this question, we conduct GenAP on the following settings: CodeBERT on the defect prediction task, CodeT5 on the code summarization and translation task, and CodeT5+ on all three tasks.
	
	Table \ref{tab:related_work_comparision} shows the results of CodeBERT in defect prediction and CodeT5 in code summarization and translation. In defect prediction, it can be observed that GenAP outperforms all rest methods by an average of 2.13\%. In code summarization, GenAP surpasses the result of the manual prompt, while all other automatic prompt design methods fall short in comparison to the manual prompt. In code translation, GenAP also surpasses the manual prompt. 


 
	Table \ref{tab:comparsion_codet5+} illustrates the results of CodeT5+ in all three tasks. Notably, GenAP operates without any manual prompt initialization, while the other methods do. Impressively, GenAP still achieves the best result compared with other methods. In defect prediction, GenAP is the only auto-prompt design method that outperforms the manual prompt. Conversely, methods such as the knowledgeable verbalizer and the soft verbalizer even fall short of the baseline result achieved without any prompt. In code summarization, the experimental results are more intriguing. In particular, GenAP is the only method that surpasses the baseline. In code translation, GenAP achieves the highest performance, while the manual prompt falls short of the baseline. These experimental results prove that GenAP is scalable to LLMs. Additionally, observations in the code summarization task and code translation task may suggest that designing effective prompts for CodeT5+ in these two tasks presents a considerable challenge.

    	
		\begin{table*}[!ht]
        \centering
		\caption{Performance Evaluation: CodeBERT in defect prediction task with Accuracy metric, and CodeT5 in Code Summarization and Translation Tasks with BLEU and CodeBLEU metrics, using various prompt design methods. Note that ``-" indicates a particular method is not suitable for specific tasks.}
		\label{tab:comparsion_codet5+}
        \begin{adjustbox}{width=0.75\textwidth}
        \begin{threeparttable}
		\begin{tabular}{@{}ccccc@{}}
			\toprule
			Methods                   & Accuracy         & BLEU (Sum.) & BLEU (Trans.)    & CodeBLEU (Trans.)       \\ \midrule
			W/o Prompt                & 51.83\%          & \underline{17.09}  & \underline{81.57}  & \underline{86.08}      \\
			Manual Prompt~\cite{schick2021exploiting}            & \underline{54.14\%}          & 16.43 & 81.45 & \underline{86.08}            \\
			Soft Prompt~\cite{qin2021learning}              & 53.88\%          & 17.01    & -   & -    \\
			Soft Verbalizer~\cite{hambardzumyan2021warp}          & 44.84\%          & -   & -  & -          \\
   			Knowledgeable Verbalizer~\cite{hu2022knowledgeable}  & 49.12\%          & -    & -  & -         \\
			GrIPS~\cite{prasad2023grips}                    & 53.92\%          & -     & -    & -      \\
			GenAP (Ours)        & \textbf{54.87\%} & \textbf{17.14} & \textbf{81.68} & \textbf{86.17} \\ \bottomrule
		\end{tabular}
            \end{threeparttable}   
        \end{adjustbox}
	\end{table*}
     \begin{table}[ht]
        \centering
		\caption{Accuracy of comparing the performance of CodeBERT and CodeT5+ model on defect prediction task via GenAP and other combined methods.}
		\label{tab:combine_methods}
        \begin{adjustbox}{width=0.95\linewidth}
        \begin{threeparttable}
          \begin{tabular}{@{}c|cc@{}}
			\toprule
			Methods & CodeBERT & CodeT5+ \\ \midrule     
            Manual Prompt* & 54.10\% & \underline{54.14\%}    \\
			Soft Prompt \& Soft Verbalizer & \underline{55.93\%} &     51.24\%           \\
            GrIPS \& Knowledgeable Verbalizer & 51.93\% & 44.22\% \\
			GenAP       & \textbf{56.19\%} & \textbf{54.87\%} \\ \bottomrule
		\end{tabular} 
            \begin{tablenotes}
              \footnotesize
              \item[*] Baseline methods
            \end{tablenotes}
        \end{threeparttable}
        \end{adjustbox}
	\end{table}
 
	To further show the effectiveness of GenAP, we attempt to combine existing methods in the defect prediction task and compare them with GenAP. We can observe from table \ref{tab:combine_methods} that GenAP continues to perform stable and well. In the defect prediction task using CodeBERT, the soft prompt \& soft verbalizer method manages to demonstrate comparable performance to GenAP,  with a 0.26\% decrease in accuracy. However, when using CodeT5+, the performance of the soft prompt \& soft verbalizer method experiences a significant drop, demonstrating a 2.9\% decrease compared with Manual Prompt. In contrast, GenAP still outperforms the manual prompt with 0.73\%. This could be due to the huge parameter size of LLMs, which poses an obstacle for those ``soft" methods to learn a prompt by gradient descent. In addition, the GrIPS \& knowledgeable verbalizer method performs poorly in the defect prediction task, exhibiting a 2.17\% decrease with CodeBERT and a more substantial 9.92\% drop with CodeT5+. Those observations further prove the scalability of GenAP to LLMs.

    \begin{table}[!h]
        \centering
		\caption{Accuracy of comparing the performance of CodeBERT model on defect prediction task via GenAP and Soft Prompt \& Verbalizer. Both methods discard the manual prompt initialization.}
		\label{tab:wo_ini}
        \begin{adjustbox}{width=\linewidth}
          \begin{tabular}{@{}ccc@{}}
			\toprule
			Methods                   & Accuracy (w/o manual) & Accuracy (w manual)          \\ \midrule
			Soft Prompt \& Verbalizer  & 55.16\%    &     55.93\%        \\
			GenAP     & \textbf{55.31\%}  & \textbf{56.19\%} \\  \bottomrule
		\end{tabular}  
        \end{adjustbox}
	\end{table}


    \subsection{RQ2: Does GenAP Require a Labor-intensive Manual Prompt Design Process?}In the CodeT5+ experiments, GenAP has initially demonstrated its performance in a fully automatic prompt design scenario. GenAP surpasses all other methods in all tasks. Moreover, this is even achieved under experimental conditions that unfairly disadvantage it because GenAP doesn't use any manual prompt initialization. For this, we additionally conduct experiments involving GenAP and the second-best approach soft prompt \& verbalizer in defect prediction using CodeBERT. Both methods are initialized without any manual prompt. Upon the removal of manual prompt initialization, both methods exhibit a performance decrease, yet GenAP still performs better, as shown in Table \ref{tab:wo_ini}. Moreover, it is noteworthy that even after removing manual prompt initialization, GenAP still surpasses the performance of existing methods shown in Table \ref{tab:related_work_comparision}. This further highlights the superiority of GenAP and proves that GenAP requires no labor-intensive manual prompt design.

    \subsection{RQ3: What Is the Convergence Speed of GenAP, and What Is the Rationale of Setting the Mutation Probability to 0.4?} We adopt average iterations to obtain a good prompt to represent the convergence speed of GenAP. The average iterations to obtain a good prompt is 15 for defect prediction, 10 for code summarization, and 13 for code translation. As for the rationale behind setting the mutation probability to 0.4, we based this choice on the assumption that when using GenAP to optimize user prompts, general users may lack professional expertise in prompt design. Consequently, these prompts might have lower quality and negatively impact the performance of the final prompt. To validate this assumption, we conducted experiments on various mutation probability settings. Initially, we set the probability to 0.2, following the approach of Sun \textit{et al.}~\cite{sun2020automatically} and find that the convergence is reached at a very early stage (about 6 iterations) when using a manual prompt as initialization. After 20 iterations are finished, we find almost all those auto-designed prompts mimic the initial manual prompt and their performance is nearly the same as the initial. This is contrary to our expectations that GenAP should optimize the initial prompt rather than mimic it. We also do experiments on 0.8 mutation probability and find the convergence is hard to reach, where the prompt design process is similar to random search. In summary, adjusting the mutation probability to 0.4 is beneficial to optimize the prompt.

    \begin{figure}[!h]
     \centering
      \includegraphics[width=\linewidth]{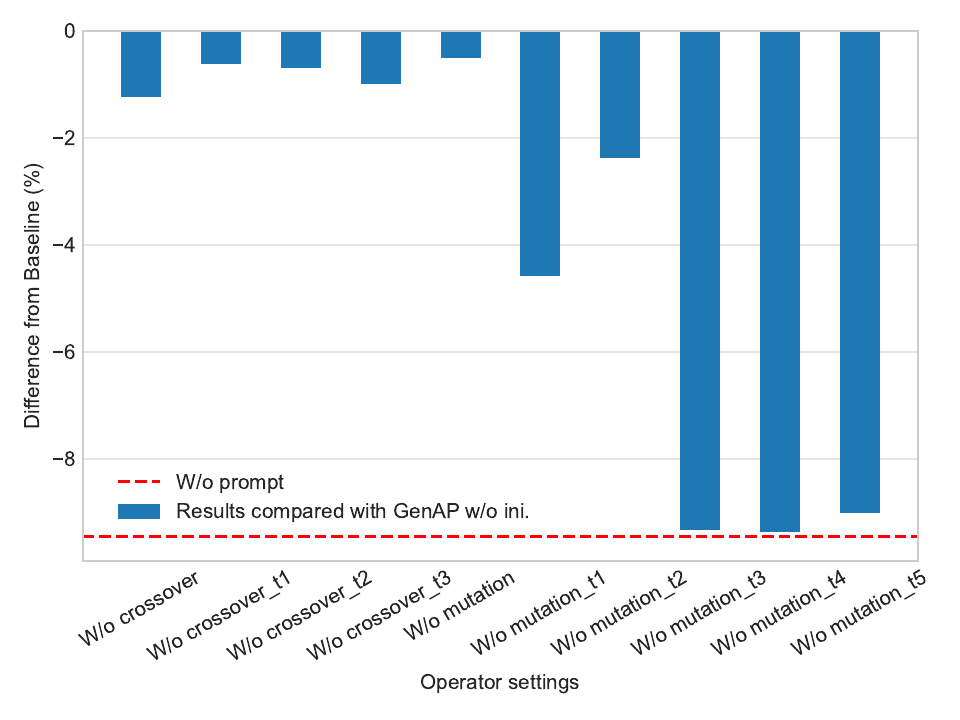}
      \caption{Ablation results with different operator settings. Note that axis 0 denotes the result of GAP w/o manual prompt initialization.}
      \label{fig:fig4}
    \end{figure}

	\subsection{RQ4: How Is the Effectiveness of each Operator Designed in GenAP?} We conduct ablation experiments on various operator settings in the defect prediction task, as the operators utilized by the verbalizer are not activated for the code summarization and code translation task. For this purpose, we systematically explore the impact of each operator while keeping the remaining operators set to their default values.
	
	As depicted in Fig.~\ref{fig:fig4}, it becomes evident that the performance of GenAP experiences a decline upon the removal of any of the operators. Notably, when the Type 3, Type 4, and Type 5 mutation operators are removed individually, the performance of GenAP dramatically drops and is even close to the results achieved without any prompts. This observation potentially underscores the significance of both the prompt template order (within the framework of the proposed encoding strategy) and the design of the verbalizer.
	

     \begin{figure}[!h]
     \centering
      \includegraphics[width=\linewidth]{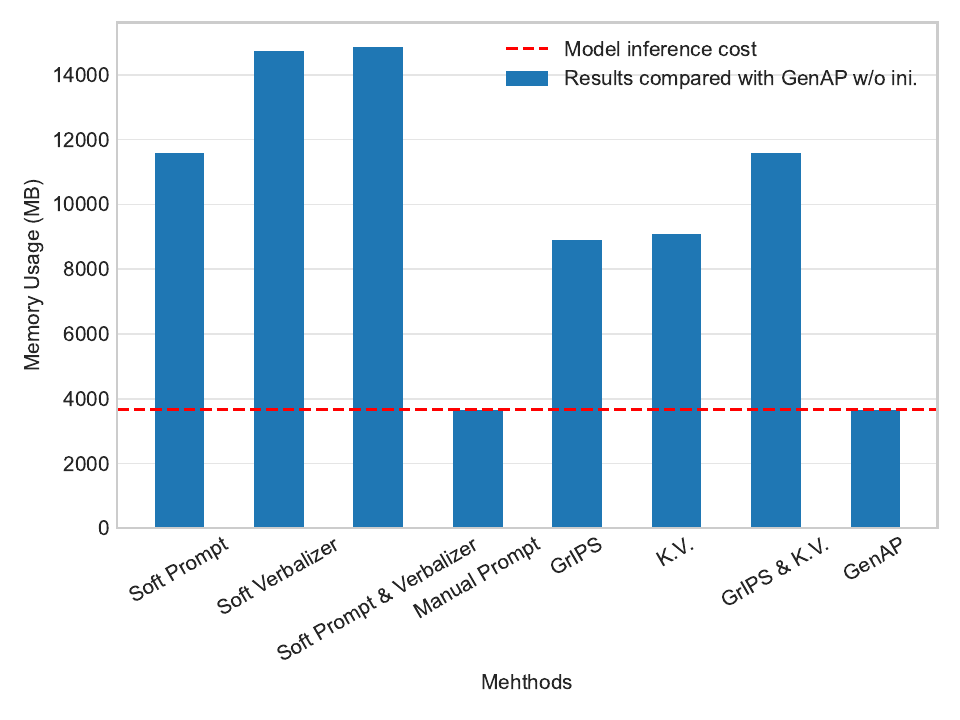}
         \caption{Memory Usage with prompt design methods when applying CodeBERT to defect prediction.}
      \label{fig:fig5}
    \end{figure}

	\subsection{RQ5: What Is the Computational Cost for GenAP?} As previously discussed, the methods that rely on discrete prompts, such as GPS, require high computational resources. To determine if GenAP faces the same challenge, we conducted experiments to observe the memory consumption of each method when applying CodeBERT to the defect prediction task. 
 
     The results are presented in Fig.~\ref{fig:fig5}. As can be observed,  GenAP and the manual prompt consume the lowest memory, which corresponds to the computational cost of model inference alone. Consequently, both methods do not impose any additional memory cost. In contrast, the methods tuning soft prompts or verbalizers demonstrate three times higher memory usage compared to GenAP. In addition, other automatic prompt design methods also exhibit considerable memory consumption when compared to GenAP. These experimental findings confirm the low computational cost characteristic of GenAP.

        \section{Discussion}
        \subsection{Design Pattern of Prompt}
        Through the experiments, we have observed several intriguing phenomena:
        \subsubsection{The Design of Verbalizers Seems to Be More Important than the Design of Prompt Templates} As shown in Table \ref{tab:pl_defect}, when using the same template with different verbalizers, the accuracy ranges from 47.44\% to 54.10\%, employing different templates yet with the same verbalizer yields accuracies from 45.86\% to 47.95\%. These findings suggest that modifying verbalizers has a greater impact on CodeBERT's performance compared to modifying prompt templates in defect prediction tasks. Moreover, results in Table \ref{tab:related_work_comparision} show methods adjusting verbalizers like soft verbalizer, soft prompt \& verbalizer, and GenAP have better performance compared to the methods adjusting prompt templates. As shown in Table \ref{tab:comparsion_codet5+}, although the methods focusing on verbalizer adjustments do not consistently outperform those focusing on prompt templates, their performance still exhibits significant variability, ranging from 44.84\% to 54.87\%. This may imply the challenge of designing a suitable verbalizer. Additionally, the results illustrated in Fig \ref{fig:fig4} reveal a sharp performance drop for GenAP when mutation operators of the verbalizer (Type 4 and Type 5) are not utilized. The automatic design of verbalizers can be viewed as the main contributor to the superior performance of GenAP.

        \subsubsection{The Prompt Template Order (within the Framework of the Proposed Encoding Strategy) Appears to Be a Crucial Factor in the Prompt Template Design} This observation is supported by the findings in Fig \ref{fig:fig4}, which demonstrates that the removal of mutation operator Type 3 has a more pronounced impact on model performance compared to mutation operator Type 1 and Type 2. We also find that the better prompt template designed by GenAP tends to have an order like ``[T] [T] [T]\textit{ \textless code\textgreater\ \textless mask\textgreater}", where [T] denotes a natural language token. This could be an interesting future work: exploring the optimal sequence of prompt templates and the underlying rationale.

        \subsubsection{Semantic Richness Is Not a Must for Good Prompts} This could be very counterintuitive since we view prompts as additional task-specific knowledge. Soft Prompt, needless to say, is not semantically rich to humans since they are even not interpretable. The prompts designed by GenAP are also not semantically rich to humans (e.g. \textit{``\textless code\textgreater\ proximity \textless mask\textgreater."}). Some work in NLP~\cite{webson2022prompt,prasad2023grips} has also observed this phenomenon. In the next subsection, we will further discuss this phenomenon by examining specific cases.

    \begin{figure}[h]
		\includegraphics[width=\linewidth]{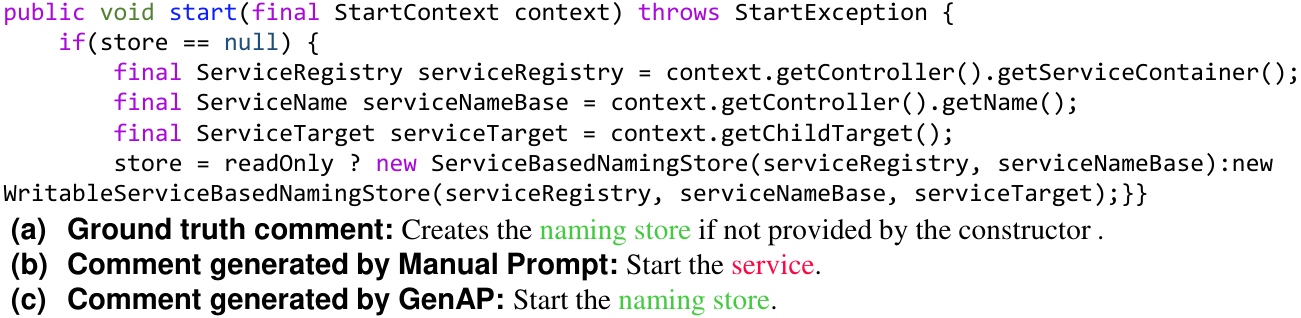}
		\caption{Case study on the code summarization task, where the pre-trained model is CodeT5.}
		\label{fig:fig6}
	\end{figure}

    \subsection{Case Study}
        To qualitatively compare GenAP with the manual prompt, we provide additional case studies. 
        
        The case in Fig.~\ref{fig:fig6} shows the Java code snippet with comments generated by CodeT5 with Manual Prompt and GenAP. We utilize the top-performing prompt ``\textit{Code \textless code\textgreater\ Summarization \textless mask\textgreater.}" in Table~\ref{tab:pl_summarization} for Manual Prompt. The prompt auto-designed by GenAP is ``\textit{ude layered sponsoring \textless code\textgreater\ \textless mask\textgreater.}".  From the case, we can observe that CodeT5 with Manual Prompt is misled by the word ``service" in the code snippet and fails to capture the main functionality ``naming store". On the contrary, CodeT5 with GenAP accurately captures ``naming store".
    \begin{figure}[h]
    		\includegraphics[width=\linewidth]{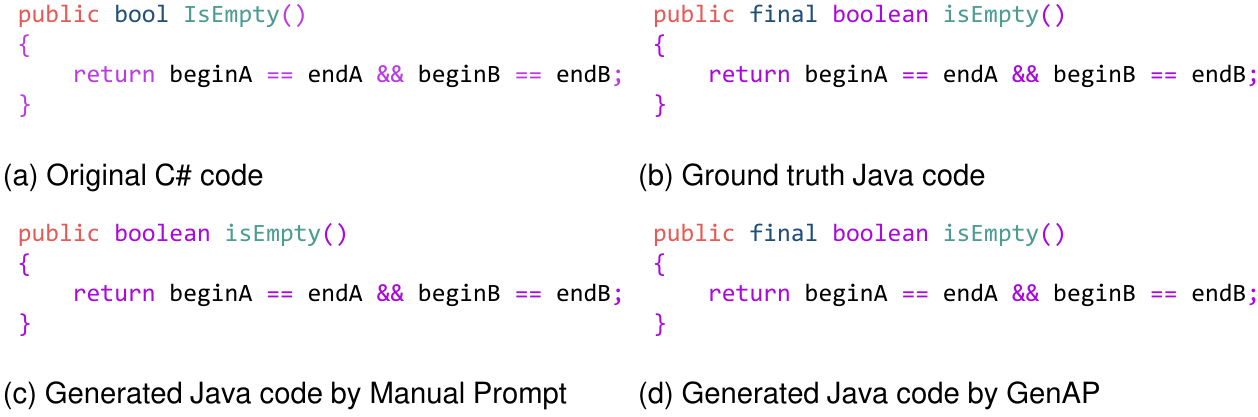}
    		\caption{Case study on the code translation task, where the pre-trained model is CodeT5.}
    		\label{fig:fig7}
	\end{figure}

        The case in Fig.~\ref{fig:fig7} shows the translated Java code snippet generated by CodeT5 with Manual Prompt and GenAP. We utilize the top-performing prompt ``\textit{Translate \textless code\textgreater\ into java \textless mask\textgreater.}" in Table~\ref{tab:pl_translation} for Manual Prompt. The prompt auto-designed by GenAP is ``\textit{IST \textless code\textgreater\ sofa Class \textless mask\textgreater.}". In this case, we can observe that the functionality of the original C\# code Fig.~\ref{fig:fig7}(a) is to check if two pairs of variables are equal. Both the Java code translated by Manual Prompt Fig.~\ref{fig:fig7}(c) and GenAP Fig.~\ref{fig:fig7}(d) successfully replicate the main functionality of the original C\# code. However, the code translated by Manual Prompt overlooks the implied meaning in the original C\# code that this method cannot be overridden by subclasses. The code translated by Manual Prompt uses \textit{``public boolean isEmpty()"}, which means this method can be overridden by subclasses. CodeT5 with GenAP, on the other hand, successfully captures the implied meaning and declares this method cannot be overridden by using \textit{``public final boolean isEmpty()"}. In addition, CodeT5 with GenAP achieves the generation of identical Java code with the ground truth Fig.~\ref{fig:fig7}(b).
        
        An even more intriguing fact is that in both two cases, the prompts auto-designed by GenAP seem semantically poor to humans but perform better, which is consistent with Section VI-A-3) ``\textit{Semantic richness is not a must for good prompts}". We believe this is a result of the inherent difference between programming languages (source code) and natural language, akin to two distinct languages like Chinese and English. Just as communication challenges between speakers of Chinese and English, natural language prompts can be perplexing for code intelligence PLMs. Therefore, auto-prompt design becomes even more crucial. Crafting a semantically rich prompt for humans may be effortless for software engineering experts. However, designing an understandable prompt for code intelligence PLMs is notably challenging.
        
	\subsection{Threats to Validity}
	Major threats to validate are illustrated as follows:
	
	\subsubsection{Limited Datasets} The experiment results are based on a limited number of datasets for each code intelligence task, which may introduce bias to the results. To address this concern, we have selected the most commonly used datasets for each task and fixed the seeds. Moreover, we intend to gather more datasets in the future for a better evaluation of prompt learning and GenAP.
	
	\subsubsection{Limited Downstream Tasks} The experiments are conducted on three canonical tasks in code intelligence, including one understanding task and two generation tasks. Although these tasks are the representative ones in code intelligence, there are many other tasks like code search~\cite{gu2018deep,cambronero2019deep} and code generation~\cite{svyatkovskiy2020intellicode,li2022competition}. We anticipate that similar observations can be made across these tasks, as they can all be categorized as either understanding or generation tasks for source code.
	
	\subsubsection{Algorithm Convergence Criterion} 
        In the experiments, we noticed varying convergence speeds among different PLMs and tasks. The current fixed convergence criterion used may prolong the runtime of GenAP. Creating an adaptive criterion for diverse PLMs and tasks is an exciting future research.

    \subsubsection{Word Pool Construction} GenAP requires the word pool for initialization. Its quality will inevitably affect the performance of GenAP. Notably, for the experiments involving GenAP on CodeBERT, we observed a more significant performance improvement compared to others. This enhancement can be attributed to the fact that the word pool used is constructed using the vocabulary of CodeBERT. Thus, designing a stronger word pool with minimal labor and expertise will be a meaningful avenue for future research. 

    \begin{table}[!h]
        \centering
		\caption{Comparsion of the performance of Prompt learning and Instruction tuning using CodeT5 and CodeT5+ model in Code Translation task with BLEU and CodeBLEU metrics. }
		\label{tab:inst_tuning}
        \begin{adjustbox}{width=0.90\linewidth}
          \begin{tabular}{@{}cc|cc@{}}
            \toprule
            \multicolumn{2}{c|}{Mehtods}                  & BLEU  & CodeBLEU \\ \midrule
            \multirow{2}{*}{CodeT5}  & Prompt learning    & 79.85 & 85.17    \\
                                     & Instruction Tuning & 44.67 & 66.37    \\
            \multirow{2}{*}{CodeT5+} & Prompt learning    & 81.63 & 86.17    \\
                                     & Instruction Tuning & 80.56 & 85.37    \\ \bottomrule 
            \end{tabular}
        \end{adjustbox}
	\end{table}

    \subsection{Comparison with Instruction Tuning}
    Instruction tuning~\cite{wei2021finetuned} is designed for LLMs to follow instructions. It can potentially enhance the performance of LLMs but may not be suitable for these relatively small PLMs. However, in recent years, the software engineering community has witnessed the ability of these relatively small PLMs~\cite{zhou2021assessing,chen2022transferability,hadi2022effectiveness} and the experiments show these PLMs have comparable performance with LLMs in certain tasks. Conducting code summarization using CodeT5 can achieve 17.29 BLEU while CodeT5+ can only achieve 17.09 BLEU. Using a state-of-the-art code LLM (CodeT5+) is even worse than the small PLM (CodeT5) in the code summarization task. Therefore, less compatibility with small-size PLMs will be a drawback to instruction tuning. We have experimented with its compatibility with CodeT5 and CodeT5+ by adapting instruction tuning to the code translation task. The instruction used for instruction tuning is \textit{``Below is an instruction that describes a task, paired with an input that provides further context. Write a response that appropriately completes the request. Instruction: Translate the following Csharp code into java. Input:\textless code\textgreater\ Response:"}. As outlined in Table \ref{tab:inst_tuning}, we find it will lead to a 34 drop of BLEU and an 18 drop of CodeBLEU for CodeT5, as well as a 1 drop of BLEU and a 0.7 drop of CodeBLEU for CodeT5+. These results confirm that instruction tuning may not be suitable for small PLMs.

	
	\section{Related Work}
	This work is related to research on pre-trained code intelligence language models, prompt learning methods, as well as automatic prompt design methods. We summarize related works as below.
 
	\subsection{Pre-trained Code Intelligence Language Models} Code intelligence leverages artificial intelligence to help software developers improve the productivity of the development process. It supports various scenarios such as code summarization, translation, and defect prediction~\cite{zhou2019devign,li2019improving,pradel2018deepbugs,hu2018summarizing,nguyen2015divide}. Recently, driven by the success of PLMs in NLP, A surge in PLMs for programming languages(PL) emerges. CodeBERT~\cite{feng2020codebert} is one of the pioneer works. It can learn NL-PL representation via replaced token detection tasks. GraphCodeBERT~\cite{guo2020graphcodebert} leverages a data flow graph to pre-train the model and make it better understand the code structure. Apart from the aforementioned encoder-only models, PLMs with other architectures are also proposed for PL. For example, Wang \textit{et al.}~\cite{wang2021codet5} modify the T5 model and introduce CodeT5, which is an encoder-decoder architecture model. As for decoder-only models, Svyatkovskiy \textit{et al.}~\cite{svyatkovskiy2020intellicode} train GPT-2~\cite{radford2019language} on a large code corpus to solve code completion task. As the size of models continues to grow, an increasing number of larger models are being proposed. Roziere \textit{et al.}~\cite{roziere2023code} further pre-train Llama 2~\cite{touvron2023llama} on a mixed dataset of natural language and code, and propose a series of Code Llama models. Their parameter size varies from 7B to 70B. Wang \textit{et al.}~\cite{wang2023codet5+} scale up the encoder-decoder model and introduce CodeT5+ models with parameter sizes ranging from 110M to 16B. Those large PLMs are commonly referred to as code LLMs. In this paper, we investigate both code intelligence PLMs and LLMs.
	
	\subsection{Prompt Learning Methods} The concept of prompt learning is developed gradually. The use of prompt can be traced back to T5~\cite{raffel2020exploring} and GPT-3~\cite{brown2020language}. Specifically, Raffel \textit{et al.}~\cite{raffel2020exploring} add a task-specific prefix to convert all text-based
	language problems into a text-to-text format, gaining a huge performance boost. Brown \textit{et al.}~\cite{brown2020language} use prompts to give GPT-3 knowledge of studied tasks. Subsequently, prompt-based methods diverged into two factions. One faction employed prompts to facilitate better model tuning~\cite{hu2022knowledgeable,wei2021finetuned}, while the other faction dives into tuning-free prompting~\cite{winata2021language,yang2022empirical,reif2022recipe}. Prompt learning methods primarily focus on the latter faction. In code intelligence, there are works that align with the former faction, such as~\cite{fan2023dialog}, which use prompts to improve fine-tuning T5 and generate summary recommendations for annotators. Works that align with the latter faction are relatively less. For example, Liu \textit{et al.}~\cite{liu2023wants} designs a method to convert user queries to prompts, enhancing the human-computer interaction of code-generating PLMs.

        \subsection{Automatic Prompt Design Methods} Note that several representative automatic prompt design methods have been introduced thoroughly in the previous sections, here we mainly discuss the relationships with GAP3~\cite{ijcai2023p588}. GAP3 utilizes a genetic algorithm to automatically design prompts in black box scenarios where the access to PLMs is limited to a cloud API. In such situations, methods for tuning soft prompts require additional APIs for vector injections, and methods for automatically designing discrete prompts require manual prompts as initialization. It appears that GAP3 addresses these issues. However, it overlooks the automatic design of the verbalizer and the prompt template order, which we have discussed the significance of in subsection 6.1. The design of a verbalizer and prompt template order in GAP3 still requires human effort, whereas GenAP does not require any human intervention. Additionally, as described in subsection 4.1, the genetic algorithm serves as a comprehensive framework that needs to be implemented in practical applications. The implementation of GAP3 differs significantly from GenAP, and their mutation operator introduces additional computational demands.

	\section{Conclusion}
	In this paper, we conduct an empirical study on the effectiveness of prompt learning in three canonical code intelligence tasks and find it largely depends on the manual design of prompts. Moreover, we assess popular approaches for automatically designing prompts in NLP and discover that methods of tuning soft prompts suffer from a lack of interpretability, and are not applicable when gradients are inaccessible. Additionally, approaches that involve tuning discrete prompts face challenges such as high computational demands and restricted applicability. Motivated by the empirical study, we introduce GenAP, a gradient-free and cost-effective automatic prompt design method. The experiments demonstrate that prompts designed by GenAP outperform meticulously hand-crafted ones, effectively automating the process of designing prompts.




\bibliographystyle{IEEEtran}
\bibliography{reference}

\vfill

\end{document}